\documentclass[letter]{aa}

\usepackage{amsmath,amssymb,amsfonts}
\usepackage{physics}
\usepackage{bm}
\usepackage{newtxtext,newtxmath}
\usepackage{graphicx}

\usepackage{times}

\usepackage[normalem]{ulem}
\usepackage{cancel}

\begin{document}

\title{Spectral decoherence without depolarization in curvature radiation}

\titlerunning{Spectral decoherence in curvature radiation}

\author{A.~G.~Tevzadze\inst{1,2} \and N.~L.~Shatashvili\inst{2,3}}

\institute{
Evgeni Kharadze Georgian National Astrophysical Observatory, Abastumani 0301, Georgia
\and
Department of Physics, Faculty of Exact and Natural Sciences, Ivane Javakhishvili Tbilisi State University, Tbilisi 0179, Georgia
\and
Andronikashvili Institute of Physics, TSU, Tbilisi 0177, Georgia
}

%\\ \email{your.email@domain}

\date{\today}

\abstract
% context heading (optional)
{Fast radio bursts often exhibit strong linear polarization together with pronounced spectral structure. Yet in coherent curvature radiation, spectral decoherence and depolarization are not necessarily simultaneous.}
% aims heading (mandatory)
{We investigate spectral and polarization coherence in curvature radiation from extended ultrarelativistic sources.}
% methods heading (mandatory)
{Using the coherency matrix formalism together with an asymptotic phase expansion for a uniformly emitting extended source, we derive the spectral and polarization coherence properties of curvature radiation.}
% results heading (mandatory)
{We show that spectral decoherence and depolarization are governed by distinct physical conditions and therefore develop on separate scales. Retardation phase variations suppress spectral coherence, whereas polarization remains largely preserved across the relativistic beaming cone, naturally producing a broad polarized but spectrally decoherent regime whose
\textcolor{black}{relative} extent increases toward higher frequencies and larger Lorentz factors.}
% conclusions heading (optional)
{Spectral decoherence without strong depolarization naturally arises in ultrarelativistic curvature radiation and may explain highly polarized fast radio bursts with strong spectral modulation or narrow band structure.}

\maketitle

\section{Introduction}

Coherent radiation from relativistic charge distributions is believed to play an important role in the observed emission from a wide range of astrophysical environments. Fast radio bursts (FRBs), whose extreme brightness temperatures require emission processes with a high degree of phase coherence among radiating particles, provide a powerful observational testbed for coherent emission mechanisms (see \cite{Katz2014,Melrose2017,Kumar2017,Lu2018,Cordes2019,Zhang2020}). 

The key observational signatures of coherent radiation are its spectral and polarization properties, both of which encode the coherence structure of the emitting region (see \cite{Petroff2019,Zhang2023}). Curvature radiation from relativistic charges moving along magnetic field lines is widely regarded as one of the leading mechanisms responsible for such emission (see \cite{Kumar2017,Lu2018,Yang2018,Ghisellini2018}). The interplay between spectral and polarization coherence is therefore central to the interpretation of coherent curvature radiation in ultrarelativistic astrophysical sources (see \cite{Wang2022,Liu2023}).

Interestingly, FRBs frequently exhibit strong linear polarization together with pronounced spectral modulation, including fine spectral structure and frequency-dependent intensity variations (see \cite{Petroff2019,Hessels2019}). In coherent radiation models, such spectral sub-structure is often interpreted as evidence for partial loss of phase coherence within the emitting region, while the observed emission commonly remains highly polarized. Such coexistence of strong polarization and pronounced spectral structure is commonly attributed to additional propagation, scattering, or plasma effects acting on the emitted radiation (see \cite{Lyubarsky2021}). This raises the question of whether highly polarized yet spectrally decoherent emission can arise naturally from curvature radiation itself, or whether additional physical mechanisms are required to explain the different evolution of spectral and polarization coherence.

Many treatments of curvature radiation formulate coherence in terms of compact charge bunches or solitons smaller than the radiation wavelength (see \cite{Ruderman1975,Melikidze2000,Yang2018}). However, spectral phase coherence and polarization coherence are governed by different physical mechanisms: phase variations across the source suppress spectral coherence, whereas polarization coherence depends on variations of the local polarization basis. This distinction may be important for highly polarized FRBs with fine spectral structure and frequency decorrelation.

In this work, we develop a coherency matrix formulation of curvature radiation from an extended ultrarelativistic source and derive distinct characteristic scales governing spectral decoherence and depolarization. We show that spectral decoherence without depolarization is a robust property of partially coherent curvature radiation and may provide a natural explanation for highly polarized FRBs exhibiting strong spectral modulation or narrow band structure. 

In Sect.~2, we derive the corresponding coherence and depolarization scales within a unified coherency-matrix formalism and discuss their implications for FRBs. We summarize our results in Sect.~3.
{\color{black} Additional discussion of the limits of validity of the present formalism, including line-of-sight (LOS) effects, is provided in the online material.
}

\section{Coherency Matrix Formalism}

Let us consider a spatially extended emitting region of size $L$, containing relativistic charged particles moving along a curved magnetic field lines.
Let us consider the vicinity of the point where the particle velocity is directed along the line of sight.
Let $s = vt$ denote the arc-length coordinate along the trajectory, where $v \simeq c$.
Using the standard far field superposition of radiation amplitudes (\cite{Landau1975}, Jackson 1998), the electric field can be calculated as follows:
\begin{equation}
\mathbf{E}(\omega,L) = \int_{-L/2}^{L/2}
\mathbf{e}(\omega,s) \, \sigma(s) \, e^{i\phi(\omega,s)} \, ds ~,
\label{eq:E}
\end{equation}
where $\omega$ is the radiation angular frequency, $\sigma(s)$ is the net emitting charge density along the trajectory, $\phi(\omega,s)$ is the radiation phase, and $\mathbf{e}(\omega,s)$ is the local polarization unit vector satisfying:
$ \mathbf{e}\cdot\mathbf{e}^{\ast}=1 $.
The total net charge is then:
\begin{equation}
\mathcal{Q}  = \int_{-L/2}^{L/2}  \sigma(s)\,ds ~.
\label{eq:Q}
\end{equation}

The coherence and polarization properties of the radiation are described by the coherency matrix
(see \cite{Goodman1985,Born1999}):
\begin{equation}
J_{ij}(\omega,L) = E_i(\omega,L) E_j^\ast(\omega,L) ~.
\end{equation}
where $i,j$ denote two orthogonal components transverse to the line of sight. 
Substituting the electric field integral gives:
\begin{eqnarray}
J_{ij}(\omega,L) &=& \iint_{-L/2}^{L/2}  \sigma(s)\sigma(s') 
P_{ij}(\omega,s,s') e^{i\Phi(\omega,s,s')} \,{\rm d} s \, {\rm d} s' \,, \nonumber \\ [4pt]
P_{ij}(\omega,s,s') &=& {e}_i(\omega,s) \, {e}_j^\ast(\omega,s') ~, \label{eq:Jij} \\ [4pt]
\Phi(\omega,s,s') &=& \phi(\omega,s)-\phi(\omega,s') ~. \nonumber 
\end{eqnarray}
This formulation makes explicit the distinction between spectral decoherence and depolarization: the phase difference $\Phi$ governs the interference between radiation emitted from different regions of the trajectory, while the kernel $P_{ij}$ encodes correlations of the local polarization basis across the source.

In the orthogonal polarization basis $(1,2)$, the Stokes parameters are obtained from the coherency matrix, with the total spectral intensity given by its trace:
\begin{equation}
I = {\rm Tr} \, J = J_{11}+J_{22} ~, \label{eq:I}
\end{equation}
while the remaining Stokes parameters are give by:
\begin{align}
Q = J_{11}-J_{22} ~,~~~ U = J_{12}+J_{21} ~,~~~ V = i\left(J_{12}-J_{21}\right) \label{eq:V} ~.
\end{align}

The spectral coherence degree may be defined as the ratio of the spectral intensity in the partially coherent case to that in the fully coherent limit, thereby equivalently characterizing the coherence efficiency of the radiation:
\begin{equation}
\eta_s(\omega,L) = {I(\omega,L)}/{I_0} ~.
\end{equation}
In the fully coherent limit, the radiation phase remains constant across
the source, $\phi(\omega,s)=\mathrm{const}$, and the polarization
basis is fixed along the trajectory,
$\mathbf e(\omega,s)=\mathbf e(\omega)$.
Thus, in fully coherent limit the coherency matrix satisfies
$I_0 = {\rm Tr} \, J = \mathcal{Q}^2 $.
Thus, the spectral coherence degree can be written as:
\begin{equation}
\eta_s(\omega,L) = {J_{ii}(\omega,L)}/{\mathcal{Q}^2} ~.
\end{equation}

The polarization coherence degree, equivalent to the usual polarization
fraction, is given by
\begin{equation}
\eta_p(\omega,L) = {\sqrt{Q^2+U^2+V^2}}/{I} ~,
\end{equation}
with $V=0$ for linearly polarized radiation.

Hence, the parameters $\eta_s(\omega,L)$ and $\eta_p(\omega,L)$ describe the spectral and polarization coherence properties of curvature radiation across different frequencies and source scales, providing a unified framework for the description of the spectral decoherence and depolarization of curvature radiation.

%To evaluate these coherence parameters, we now construct the
%radiation phase and local polarization basis.

\subsection{Radiation Phase}

For a relativistic charged particle moving along a curved trajectory, the radiation phase is determined by the retarded time structure of the Liénard--Wiechert potentials,
\begin{equation}
\phi(\omega,t) = \omega \left(\frac{s}{v} - \frac{\bm{n}\cdot \bm{r}(s)}{c}\right),
\label{eq:phase_LW}
\end{equation}
where $\bm{r}(s)$ is the particle position, $\bm{n}$ is a unit vector pointing toward the observer, and $c$ is the speed of light.

We define $\rho$ as the local radius of curvature and choose coordinates such that $\bm{n}$ lies along the tangent direction at this point. This motivates the definition of a characteristic geometric coherence scale,
\begin{equation}
s_\star = {\rho}/{\gamma},
\end{equation}
corresponding to the arc length over which the direction of motion changes by the beaming angle $1/\gamma$.
Hence, it is convenient to introduce a dimensionless coordinate along the trajectory $\xi$ and the corresponding dimensionless source size $\ell$ as follows:
\begin{equation}
\xi \equiv {s}/{s_\star} ~,~~~~ \ell \equiv {L / s_\star} ~,~~~~ |\xi| \leq {\ell / 2} ~.
\end{equation}
Over the radiation formation region, the trajectory may be approximated by its local osculating circle. Introducing the angular displacement ${\color{black}\theta} \equiv {s}/{\rho} = {\xi}/{\gamma}$,
to leading order in the local circular approximation, the projection of the position vector along the line of sight is: $ \bm n \cdot \bm r(s) \simeq \rho \sin {\color{black}\theta} $.
Hence, the radiation phase can be derived as follows:
\begin{equation}
\phi(\xi) = {3 \over 2} \gamma^3 \left(\frac{\omega}{\omega_c} \right) \left[ {\xi \over \beta \gamma} - \sin \left( {\xi \over \gamma} \right) \right] ~,
\end{equation}
where $\beta = v/c$ and we have introduced the standard characteristic frequency of synchrotron radiation (see \cite{Jackson1998}):
\begin{equation}
\omega_c = \frac{3}{2}\frac{\gamma^3 c}{\rho} ~.
\end{equation}

We derive the asymptotic expansion of the radiation phase in the $|\xi / \gamma| \ll 1$ limit. Using the the ultra-relativistic approximation ${1 / \beta} \simeq 1 + {1 / 2 \gamma^2}$, together with Taylor expansion of the $\sin$ function, we obtain:
\begin{equation}
\phi = \phi_1 + \phi_3 + \cdots ~,
\label{eq:phase_series}
\end{equation}
where
\begin{equation}
\phi_1(\omega,\xi) = {3 \over 4} \left({\omega \over \omega_c}\right) \xi ~,~~~
\phi_3(\omega,\xi) = {1 \over 4} \left({\omega \over \omega_c}\right) \xi^3 ~, \label{eq:phi13} 
\end{equation}
with next term $\phi_5(\omega,\xi) \propto (\xi/\gamma)^2 \xi^3$. 
Interestingly, in the ultra-relativistic limit the phase structure of the radiation field exhibits two distinct characteristic scales. The fully coherent regime corresponds to $\xi \ll 1$, for which phase variations are negligible and the radiation adds coherently, while the local curvature expansion remains valid as long as $\xi/\gamma \ll 1$. This naturally reveals a broad intermediate domain, $1 \lesssim \xi \ll \gamma$, in which the radiation is already partially spectrally decoherent while the local curvature expansion and polarization coherence remain largely preserved.

\subsection{Depolarization}

For simplicity, we adopt a uniform net charge distribution within the emitting region, thereby isolating the intrinsic geometric effects of curvature radiation without additional plasma or propagation effects:
\begin{equation}
\sigma(s) = {\mathcal{Q}}/{L} = {\mathcal{Q}}/{(s_\star \ell)} ~.
\end{equation}
Hence, using the dimensionless coordinate $\xi$, the coherency matrix introduced in Eq.~(\ref{eq:Jij}) may be rewritten as:
\begin{equation}
J_{ij}(\omega,\ell) = \frac{\mathcal{Q}^2}{\ell^2} \iint_{-\ell/2}^{\ell/2}
P_{ij}(\omega,\xi,\xi') e^{i\Phi(\omega,\xi,\xi')} \,{\rm d}\xi\,{\rm d}\xi' ~.
\end{equation}
%We use this minimal uniform source model to demonstrate that the effect arises intrinsically, without %requiring additional plasma or propagation effects.

%Depolarization is governed by variations of the local polarization direction, encoded in the %polarization kernel $P_{ij}$. 
To isolate polarization coherence from spectral decoherence, we average over the rapidly oscillating phase factor $\Phi$. In the decoherent regime, phase oscillations suppress contributions from $\xi \neq \xi'$, so that the coherency matrix becomes approximately local and the interference terms vanish, leaving only the polarization component of the coherency matrix:
\begin{equation}
\left\langle J_{ij}(\omega,\ell) \right\rangle = \frac{\mathcal{Q}^2}{\ell} \int_{-\ell/2}^{\ell/2}
P_{ij}(\omega,\xi,\xi) \,{\rm d}\xi  ~.
\end{equation}

In general, the local polarization vector may depend on frequency, $\mathbf e=\mathbf e(\omega,\xi)$, owing to plasma propagation effects as well as the intrinsic frequency dependence of the emission geometry and magnetic field structure. In the present work, however, we focus on the purely geometric contribution to depolarization and therefore neglect propagation-induced frequency dependence of the polarization basis: $ \mathbf e(\omega,\xi)\simeq \mathbf e(\xi)$. In this case polarization basis is
defined by the local polarization angle $\chi(\xi)$ across the emitting region:
\begin{equation}
\mathbf e(\xi) = 
\begin{pmatrix}
\cos\chi(\xi) \\ \sin\chi(\xi) 
\end{pmatrix} ~.
\end{equation}
The corresponding polarization kernel is reduced to:
\begin{equation}
P_{ij}(\xi) = e_i(\xi)e_j^\ast(\xi) ~.
\end{equation}
The Stokes parameters from Eqs. (\ref{eq:I}-\ref{eq:V}) are reduced to $I = \mathcal{Q}^2$
and the following combination:
\begin{equation}
Q + i U = \frac{\mathcal Q^2}{\ell} \int_{-\ell/2}^{\ell/2} e^{2i\chi(\xi)} \,d\xi ~.
\end{equation}
Hence, the polarization coherence degree becomes:
\begin{equation}
\eta_P(\ell) = 
\left| \frac{1}{\ell} \int_{-\ell/2}^{\ell/2} e^{2i\chi(\xi)} \,d\xi \right| ~,
\end{equation}
which measures polarization alignment across the source.

In ultra-relativistic regime, the local polarization direction is expected to vary across the emitting region, following the local curvature of magnetic field lines. 
Hence, Eq.~(23) gives the depolarization degree:
\begin{equation}
\eta_p(\ell) = \left| \frac{\sin(\ell/\gamma)}{\ell/\gamma} \right| ~.
\end{equation}
Thus, and significant depolarization develops only when the emitting region reaches the characteristic depolarization scale,
\begin{equation}
\ell_p \sim \gamma .
\end{equation}
This corresponds to order-unity variations of the polarization direction across the relativistic beaming cone.

\subsection{Spectral Decoherence}

Having isolated the polarization coherence properties through phase averaging, we now assume an approximately uniform polarization structure across the emitting region and focus on decoherence arising from phase variations. For a normalized polarization state with negligible polarization variations across the emitting region, the polarization kernel reduces to $ P_{ii}^{(0)} = 1 $, 
so that the spectral coherence degree is governed entirely by the phase structure of the radiation field:
\begin{equation}
J_{ii}(\omega,\ell) = \frac{\mathcal Q^2}{\ell^2}
\iint_{-\ell/2}^{\ell/2} e^{i\Phi(\omega,\xi,\xi')} \,d\xi\,d\xi' ~.
\end{equation}
Since the phase difference $\Phi$ depends only on the separation between emitting points, the exponential phase factor splits into independent terms.
The double integral therefore factorizes into an integral and its complex conjugate, yielding the squared modulus of a single phase integral. Neglecting $\phi_5$ and higher-order terms (see Eqs. 15-16), and in the $\xi/\gamma \ll 1$ limit, one obtains:
\begin{equation}
\eta_s(\omega,\ell) = \frac{1}{\ell^2} \left| \int_{-\ell/2}^{\ell/2} \exp \left[ i(\phi_1(\omega,\xi) + \phi_3(\omega,\xi)) \right] \, d\xi \right|^2 \, .
\end{equation}

To characterize the onset of spectral decoherence, we introduce the characteristic decoherence scale $\ell_s$, defined by the condition that the accumulated phase becomes of order unity,
\begin{equation}
|\phi(\omega,\ell_s)| \simeq 1 ~.
\end{equation}
Hence, using Eqs.~(15)–(16), we may derive the decoherence scale as the real positive root of the equation:
\begin{equation}
\ell_s^3  + 3 \ell_s  = 4 \left( {\omega / \omega_c} \right)^{-1} ~.
\end{equation}
Since curvature radiation is exponentially suppressed at frequencies
$\omega \gg \omega_c$, the physically relevant decoherence scale is given by
\begin{equation}
\ell_s(\omega) \simeq \left(4\omega_c/\omega\right)^{1/3} ~,~~~~
\omega \lesssim \omega_c ~.
\end{equation}

An exact solution for the spectral coherence rate at arbitrary spatial extent
of the emitting region, $\ell$, can be obtained by numerical integration of
Eq.~(28). However, it is instructive to consider the limit
$\ell \ll \ell_s$, in which spectral coherence remains nearly preserved:
\begin{equation}
\eta_s(\omega,\ell) \simeq
1 - \frac{3}{64}
\left( \frac{\omega}{\omega_c} \right)^2
\ell^2 ~,
\end{equation}
demonstrating that coherence decreases more slowly at lower frequencies.

\subsection{Decoherence without Depolarization}

Spectral coherence is governed by accumulated retardation phase differences, which grow rapidly with the spatial extent of the source. In contrast, polarization coherence depends only on the comparatively slow variation of the local polarization basis across the relativistic beaming cone.

As a result, spectral decoherence generally develops much earlier than significant depolarization. The corresponding coherence scales satisfy the hierarchy
\begin{equation}
\ell_s \ll \ell_p \sim \gamma ,
\end{equation}
revealing a broad intermediate regime in which curvature radiation becomes spectrally decoherent while remaining highly polarized.
The relative extent of this intermediate regime is characterized by the scale separation
${\ell_p}/{\ell_s}$.
Using Eqs.~(26) and (31), one obtains:
\begin{equation}
\frac{\ell_p}{\ell_s} \simeq \gamma \left( \frac{\omega}{4\omega_c} \right)^{1/3}, \qquad
\omega < \omega_c ~.
\end{equation}
Thus, the separation between the depolarization and spectral decoherence scales increases toward larger Lorentz factors and higher observing frequencies. In the ultrarelativistic regime, this naturally produces a broad spectral domain in which curvature radiation remains highly polarized despite substantial spectral decoherence.

Figure 1 illustrates the dependence of the spectral coherence $\eta_s$ and polarization $\eta_p$ degrees on the spatial extent of the emitting region $\ell$. The figure clearly demonstrates the existence of the intermediate decoherence regime, where the radiation loses spectral coherence while largely preserving its polarization.

These results suggest that spectral decoherence without significant depolarization is an intrinsic property of curvature radiation in the ultrarelativistic regime. Consequently, highly polarized emission with fine spectral structure may naturally emerge from curvature radiation itself, without requiring additional physical mechanisms to produce spectral decoherence.

The present analysis further predicts that the extent of the intermediate regime increases with the Lorentz factor. Sources with larger $\gamma$ are therefore expected to exhibit a broader frequency range over which the radiation remains highly polarized despite substantial spectral decoherence. This provides a potentially observable signature of the coherence properties of curvature radiation in FRBs and related compact astrophysical sources.

\begin{figure}
\centering
\includegraphics[width=0.99 \columnwidth]{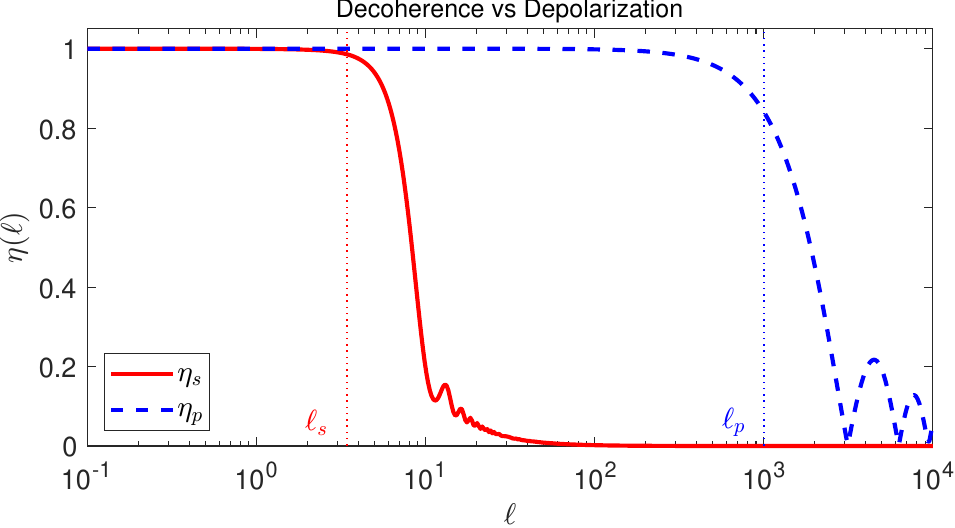}
\caption{
Spectral decoherence and depolarization degrees as functions of the source size $\ell$ for $\gamma=10^4$ and $\omega / \omega_c = 0.1$.
The spectral coherence degree $\eta_s$ decreases rapidly once the source size exceeds the decoherence scale $\ell_s$, while the polarization coherence degree $\eta_p$ remains close to unity up to the much larger depolarization scale $\ell_p$.
This hierarchy naturally produces a broad intermediate regime in which the radiation is partially spectrally decoherent while remaining highly polarized.
}
\label{fig:decoherence}
\end{figure}

\section{Summary}

We have shown that spectral decoherence and depolarization in curvature radiation are governed by distinct physical mechanisms and therefore develop on separate characteristic scales. In the ultrarelativistic regime, the spectral decoherence scale remains much smaller than the depolarization scale, naturally producing a broad intermediate regime in which the radiation becomes partially spectrally decoherent while remaining highly polarized.

We show that spectral coherence is suppressed once phase variations across the emitting region become significant, whereas polarization coherence is preserved as long as the local polarization basis varies only weakly across the relativistic beaming cone. The analysis further demonstrates that the spectral decoherence scale increases toward lower frequencies, while significant depolarization develops only on much larger scales proportional to the Lorentz factor. Consequently, the polarized but spectrally decoherent regime broadens toward higher frequencies and larger Lorentz factors.

This coherence hierarchy provides a natural framework for interpreting highly polarized FRBs exhibiting strong spectral modulation or narrow-band structure (e.g., \cite{Spitler2016,Hessels2019,CHIME2019}). Interestingly, several repeating FRBs exhibit substantial spectral variability and patchy narrow-band emission while simultaneously maintaining a high degree of linear polarization (e.g., \cite{Michilli2018,Luo2020}), qualitatively consistent with the present picture in which spectral decoherence develops substantially earlier than depolarization in ultrarelativistic curvature radiation.

For coherent curvature-radiation models of FRBs, with GHz observing frequencies and Lorentz factors $\gamma \sim 10^{2}-10^{4}$, the polarized yet spectrally decoherent regime is expected to be broad. Larger Lorentz factors further extend the frequency range over which radiation remains highly polarized despite substantial spectral decoherence, providing a potential diagnostic of the coherence properties and Lorentz factors of FRB sources.
{\color{black} Still, the detailed spectral morphology is source dependent, whereas the robust prediction is the separation between the spectral decoherence and depolarization scales.}

\begin{acknowledgements}
This work was partially supported by the Shota Rustaveli National Science Foundation of Georgia under Grant Project No. FR-22-8273. 
\end{acknowledgements}

\Online 

\onecolumn

\appendix

{\color{black}

\section{Validity of the Radiation Phase Expansion and LOS Visibility}

To capture the essential physics within a minimal framework, the analysis in the main text adopts a 1D description of the emitting trajectory. The robustness of the resulting conclusions can therefore be established by examining the validity of the underlying mathematical formalism and the extent to which the finite angular visibility imposed by relativistic beaming relative to the line of sight (LOS) affects the analysis.

For the analysis of the radiation phase, we employ the asymptotic expansion of Eq.~(15) in the ultrarelativistic limit:
\begin{equation}
\phi=\phi_1+\phi_3+\phi_5+\cdots ~,
\end{equation}
and examine the relative importance of successive terms. 
The relative importance of the linear and cubic contributions is determined by:
\begin{equation}
\frac{\phi_3}{\phi_1}=\frac{1}{3}\xi^2 ~.
\end{equation}
It should be noted that the linear contribution, $\phi_1$, does not arise directly from the asymptotic expansion of the sine function. Rather, it results from the partial cancellation between the linear contributions of the first two terms in the phase, leaving a residual term of order $\gamma^{-2}$ (see Eq.~13). Consequently, the ratio $\phi_3/\phi_1$ must be evaluated separately. For the higher-order terms originating solely from the Taylor expansion of the sine function, the ratio of successive odd-order contributions is then:
\begin{equation}
\frac{\phi_n}{\phi_{n-2}} = -\frac{1}{n(n-1)} \frac{\xi^2}{\gamma^2} ~,
\qquad (n = 5, 7, \ldots) ~.
\end{equation}
The resulting phase expansion therefore has a mixed asymptotic structure. The first term, 
($\phi_1$) is a kinematic residual associated with the finite difference between $v$ and $c$, whereas the higher-order terms ($\phi_{3,5,\ldots}$) arise from the curvature of the trajectory through the Taylor expansion of the sine function. Accordingly, Eq.~(A.2) characterizes the transition from the kinematic to the curvature-dominated regime, while Eq.~(A.3) quantifies the convergence of the curvature expansion. Thus, different aspects of the radiation physics are controlled by distinct dimensionless scales: $\xi$ governs the transition between the kinematic and curvature-dominated regimes, whereas $(\xi/\gamma)^2$ controls the convergence of the curvature expansion itself.

Eqs. (A.2) and (A.3) indicate that for $\xi \gtrsim \sqrt{3}$, the radiation phase is already dominated by the cubic curvature term, while the higher order curvature corrections remain asymptotically small provided $\xi/\gamma \ll 1$. Consequently, there exists a broad ultrarelativistic domain,
\begin{equation}
1 \lesssim \xi \ll \gamma ~,
\end{equation}
in which the cubic curvature term dominates over the linear contribution, while all higher order corrections are asymptotically suppressed.

The above analysis, however, is formulated within the 1D description of the emitting trajectory and does not explicitly account for the finite angular visibility of different trajectory segments relative to the observer. In the ultrarelativistic regime, relativistic beaming may restrict the portion of the trajectory that contributes effectively to the observed radiation. It is therefore necessary to estimate the effects of LOS visibility and determine whether they impose additional constraints on the range of trajectory scales relevant to the analysis.

For a fixed LOS we choose $\xi=0$ to be the point at
which the instantaneous particle velocity is tangent to the LOS.
Trajectory segments with $\xi\neq0$ are therefore viewed at a finite
angle, since the particle velocity rotates continuously as the charge
moves along the curved trajectory.

The characteristic visibility scale can be estimated directly from the
curvature contribution to the radiation phase. 
The effective contribution is confined
to the trajectory interval over which the curvature induced phase
variation remains of order unity. Defining the corresponding
dimensionless visibility scale:
\begin{equation}
\left| \phi_3(\omega,\xi_{\rm vis}) \right| \sim 1 ~,
\end{equation}
and using cubic term from the Eq. (16) we get:
\begin{equation}
\xi_{\rm vis}(\omega) \sim 4^{1/3} \left( \frac{\omega_c}{\omega} \right)^{1/3} ~.
\end{equation}
For a trajectory of curvature radius $\rho$, the tangent direction
rotates by an angle $s/\rho$ relative to the LOS tangent at $\xi=0$.
Hence, to leading order we get:
\begin{equation}
\theta(\xi) \simeq \frac{|s|}{\rho} =
\frac{|\xi|}{\gamma} ~,
\end{equation}
The corresponding characteristic visibility
angle is therefore
\begin{equation}
\theta_{\rm vis}(\omega) \sim \frac{\xi_{\rm vis}(\omega)}{\gamma}
\sim \frac{4^{1/3}}{\gamma} \left( \frac{\omega_c}{\omega} \right)^{1/3} ~,
\end{equation}

Strictly speaking, the above argument determines the characteristic
phase formation scale, defined by the condition that the
curvature induced phase variation becomes of order unity. For curvature
radiation, however, this scale is equivalent to the 
frequency dependent angular width of the radiation
pattern obtained from the exact solution. Consequently, we use
$\theta_{\rm vis}$ and $\xi_{\rm vis}$ as the characteristic
LOS visibility scales throughout this section.
Hence, the LOS visible portion of the trajectory is determined by the condition:
\begin{equation}
\theta(\xi) \lesssim \theta_{\rm vis}(\omega) ~, 
\end{equation}
or equivalently,
\begin{equation}
|\xi| \lesssim \xi_{\rm vis}(\omega) ~.
\end{equation}
Thus, the visible trajectory interval is centered on the LOS tangent
point,
\begin{equation}
-\xi_{\rm vis}(\omega) \lesssim \xi \lesssim
\xi_{\rm vis}(\omega) ~. 
\end{equation}

The corresponding physical extent is
\begin{equation}
s_{\rm vis}(\omega) \sim \rho\theta_{\rm vis}(\omega) \sim 
\frac{\rho}{\gamma} \, \xi_{\rm vis}(\omega) ~, 
\end{equation}
so that the total LOS visible trajectory length is of order
$2s_{\rm vis}$.

An important consequence is that the explicit Lorentz factor dependence
of the angular visibility cancels when the visible trajectory is
expressed in terms of the dimensionless coordinate $\xi$. Although the
physical visibility angle decreases as $1/\gamma$, the dimensionless
coordinate $\xi$ expands the corresponding physical
trajectory interval by the same factor. Consequently, the dimensionless
visibility scale is independent of $\gamma$.

Near the critical frequency, $\omega_c$, one recovers
the familiar relativistic beaming scales:
\begin{equation}
\theta_{\rm vis} \sim \frac{1}{\gamma}~, \qquad
s_{\rm vis} \sim \frac{\rho}{\gamma}~, \qquad
\xi_{\rm vis} \sim 1 ~. 
\end{equation}
For direct comparison, at lower frequencies we get:
\begin{equation}
\theta_{\rm vis} \sim \frac{1}{\gamma} 
\left( \frac{\omega_c}{\omega} \right)^{1/3}, \qquad
s_{\rm vis} \sim \frac{\rho}{\gamma} \left( \frac{\omega_c}{\omega} \right)^{1/3}, \qquad
\xi_{\rm vis} \sim \left( \frac{\omega_c}{\omega} \right)^{1/3} ~,
\end{equation}
Hence, for $\omega<\omega_c$, we get:
\begin{equation}
\xi_{\rm vis}>1 ~,
\end{equation}
and the visible region may extend beyond the conventional trajectory
scale $|\xi| < 1$. Therefore, $|\xi|>1$ does not by itself imply that a
trajectory segment is invisible. The relevant visibility criterion is
instead frequency dependent:
\begin{equation}
|\xi| \lesssim \xi_{\rm vis}(\omega) ~. 
\end{equation}

This distinction is important for the spectral decoherence analysis. At
frequencies below $\omega_c$, the LOS visible interval may extend beyond
the onset of spectral decoherence, leaving a finite domain in which
spectrally decoherent radiation remains observable. If the
polarization coherence scale exceeds the LOS visibility scale, the
observed radiation may therefore become spectrally decoherent while
remaining strongly polarized.

\section{Validity of the Depolarization Analysis under LOS Visibility}

We now examine whether restricting the emitting trajectory to the
LOS visible region modifies the depolarization analysis.
We assume that the visible contribution is centered on $\xi=0$ and
is restricted to $ |\xi| \lesssim \xi_{\rm vis}(\omega)$.
The corresponding angular displacement of the instantaneous particle
velocity from the LOS is
\begin{equation}
\theta(\xi) \simeq \frac{|\xi|}{\gamma} ~, 
\end{equation}
so that the maximum angular variation sampled within the visible region
is $\theta_{\rm vis} = \xi_{\rm vis}/\gamma$ . 

Spectral decoherence is controlled by the
accumulated radiation phase. Denoting the characteristic spectral
coherence scale by $\ell_s$, substantial phase decorrelation begins when
the effective emitting interval satisfies
\begin{equation}
\ell_{\rm eff} \gtrsim \ell_s ~,
\end{equation}
where the LOS restricted effective extent is
\begin{equation}
\ell_{\rm eff} \sim \min\!\left[ \ell,\xi_{\rm vis}(\omega) \right] ~.
\end{equation}
Thus, spectral decoherence remains observable provided that
\begin{equation}
\ell_s \lesssim \ell_{\rm eff} \lesssim \xi_{\rm vis}.
\label{eq:visible_decoherence_condition}
\end{equation}
In the low frequency curvature radiation regime,
$\ell_s$ and $\xi_{\rm vis}$ have the same characteristic scaling,
\begin{equation}
\ell_s \sim \xi_{\rm vis} \sim \left( \frac{\omega_c}{\omega} \right)^{1/3} ~.
\end{equation}
The LOS visible interval can
therefore reach the onset of phase decorrelation rather than truncating
the emission before spectral decoherence develops.

Depolarization, by contrast, requires an appreciable rotation of the
local polarization basis across the contributing trajectory. For
curvature radiation, the polarization direction follows the local
trajectory geometry and changes by an angle of order
\begin{equation}
\Delta\chi(\xi) \sim \frac{|\xi|}{\gamma} ~.
\end{equation}
An order-unity change of the polarization basis therefore occurs only on
the much larger dimensionless scale
\begin{equation}
\ell_p
\sim
\gamma.
\end{equation}
Within the LOS visible region, the total polarization angle variation is
bounded by
\begin{equation}
\Delta\chi_{\rm vis} \sim \frac{\ell_{\rm eff}}{\gamma} \lesssim 
\frac{\xi_{\rm vis}}{\gamma} = \theta_{\rm vis} ~.
\end{equation}
Consequently, whenever
\begin{equation}
\xi_{\rm vis}
\ll
\gamma,
\label{eq:polarized_visible_condition}
\end{equation}
the polarization vectors emitted throughout the visible interval remain
nearly parallel. The off diagonal components of the coherency matrix are
then not significantly suppressed by polarization basis averaging, even
when the frequency dependent phases of different trajectory segments
have become partially decorrelated.

Combining Eqs.~(\ref{eq:visible_decoherence_condition}) and
(\ref{eq:polarized_visible_condition}), the relevant depolarization-decoherence hierarchy is
\begin{equation}
\ell_s \lesssim \ell_{\rm eff} \lesssim \xi_{\rm vis} \ll \ell_p \sim \gamma ~.
\label{eq:visibility_hierarchy}
\end{equation}
This hierarchy defines a regime in which the radiation is both visible
to the observer and spectrally decoherent, while its polarization basis
remains coherent.

The same conclusion can be expressed directly in terms of the
visibility angle. Since
\begin{equation}
\frac{\xi_{\rm vis}}{\gamma} \sim \frac{1}{\gamma}
\left( \frac{\omega_c}{\omega} \right)^{1/3} ~,
\end{equation}
the condition for negligible depolarization is
\begin{equation}
\frac{1}{\gamma} \left( \frac{\omega_c}{\omega} \right)^{1/3} \ll 1 ~,
\end{equation}
or equivalently,
\begin{equation}
\frac{\omega}{\omega_c}
\gg
\gamma^{-3}.
\end{equation}
For an ultrarelativistic source, this condition is satisfied over a broad
frequency range below $\omega_c$. 

The physical reason is that spectral decoherence and depolarization
measure different properties of the radiation field. Spectral
decoherence is produced by rapidly accumulating relative phases,
whereas depolarization requires a significant geometrical rotation of
the polarization basis. Across the LOS-visible trajectory, the phase
may vary by order unity or more while the polarization direction changes
only by the parametrically smaller angle
$\xi_{\rm vis}/\gamma\ll1$. The observed radiation can therefore remain
strongly polarized despite partial spectral decoherence.

Accordingly, introducing finite LOS visibility does not invalidate the
depolarization analysis. Rather, it selects the effective trajectory
interval contributing to the observed signal while preserving the
separation between the spectral and polarization coherence scales. In
the ultrarelativistic regime, the observer can therefore receive
spectrally decoherent radiation from the visible portion of the
trajectory without a corresponding loss of linear polarization.

}


\begin{thebibliography}{}


\bibitem[Born \& Wolf (1999)]{Born1999}
Born, M. \& Wolf, E. 1999, 
\textit{Principles of Optics}, 7th edn. 
(Cambridge: Cambridge Univ. Press)

\bibitem[CHIME/FRB Collaboration et al.(2019)]{CHIME2019}
CHIME/FRB Collaboration, Amiri, M., Andersen, B. C., et al. 2019, Nature, 566, 235

\bibitem[Cordes \& Chatterjee (2019)]{Cordes2019}
Cordes, J. M., \& Chatterjee, S. 2019,
Annu. Rev. Astron. Astrophys., 57, 417

\bibitem[Ghisellini \& Locatelli (2018)]{Ghisellini2018}
Ghisellini, G., \& Locatelli, N. 2018,
Astron. Astrophys., 613, A61

\bibitem[Goodman (1985)]{Goodman1985}
Goodman, J. W. 1985, Statistical Optics (New York: Wiley)

\bibitem[Hessels et al. (2019)]{Hessels2019}
Hessels, J.~W.~T., Spitler, L.~G., Seymour, A.~D., et al. 2019,
Astrophys. J. Lett., 876, L23

\bibitem[Jackson (1998)]{Jackson1998}
Jackson, J.~D. 1998,
\textit{Classical Electrodynamics}
(3rd edn.; New York: Wiley)

\bibitem[Katz (2014)]{Katz2014}
Katz, J.~I. 2014,
Phys.\ Rev.\ D, 89, 103009

\bibitem[Katz (2018)]{Katz2018}
Katz, J.~I. 2018,
Prog. Part. Nucl. Phys., 103, 1

\bibitem[Kumar et al. (2017)]{Kumar2017}
Kumar, P., Lu, W., \& Bhattacharya, M. 2017,
Mon.\ Not.\ R.\ Astron.\ Soc., 468, 2726

\bibitem[Landau \& Lifshitz (1975)]{Landau1975}
Landau, L. D. \& Lifshitz, E. M. 1975, 
\textit{The Classical Theory of Fields}, 
4th edn. (Oxford: Pergamon Press)

\bibitem[Liu et al. (2023)]{Liu2023}
Liu, Z.-N., Yang, Y.-P., \& Zhang, B. 2023,
Astrophys. J., 943, 72

\bibitem[Lu \& Kumar (2018)]{Lu2018}
Lu, W., \& Kumar, P. 2018,
Mon. Not. R. Astron. Soc., 477, 2470


\bibitem[Luo et al.(2020)]{Luo2020}
Luo, R., Wang, B. J., Men, Y. P., et al. 2020, Nature, 586, 693

\bibitem[Lyubarsky (2021)]{Lyubarsky2021}
Lyubarsky, Y. 2021,
Universe, 7, 56

\bibitem[Melikidze et al. (2000)]{Melikidze2000}
Melikidze, G.~I., Gil, J.~A., \& Pataraya, A.~D. 2000,
Astrophys.\ J., 544, 1081

\bibitem[Melrose (2017)]{Melrose2017}
Melrose, D.~B. 2017,
Rev.\ Mod.\ Plasma Phys., 1, 5

\bibitem[Michilli et al.(2018)]{Michilli2018}
Michilli, D., Seymour, A., Hessels, J. W. T., et al. 2018, Nature, 553, 182

\bibitem[Petroff et al. (2019)]{Petroff2019}
Petroff, E., Hessels, J.~W.~T., \& Lorimer, D.~R. 2019,
Astron. Astrophys. Rev., 27, 4

\bibitem[Ruderman \& Sutherland (1975)]{Ruderman1975}
Ruderman, M. A., \& Sutherland, P. G. 1975,
Astrophys. J., 196, 51

\bibitem[Rybicki \& Lightman (1979)]{Rybicki1979}
Rybicki, G.~B., \& Lightman, A.~P. 1979,
\textit{Radiative Processes in Astrophysics}
(New York: Wiley)


\bibitem[Spitler et al.(2016)]{Spitler2016}
Spitler, L. G., Scholz, P., Hessels, J. W. T., et al. 2016, Nature, 531, 202


\bibitem[Wang et al. (2022)]{Wang2022}
Wang, W.-Y., Xu, R., \& Chen, X. 2022,
Mon. Not. R. Astron. Soc., 517, 5080

\bibitem[Yang \& Zhang (2018)]{Yang2018}
Yang, Y.-P., \& Zhang, B. 2018,
Astrophys.\ J., 868, 31

\bibitem[Yang et al. (2020)]{Yang2020}
Yang, Y.-P., Zhu, J.-P., Zhang, B., \& Wu, X.-F. 2020,
Astrophys.\ J., 901, L13

\bibitem[Zhang (2020)]{Zhang2020}
Zhang, B. 2020,
Nature, 587, 45

\bibitem[Zhang (2023)]{Zhang2023}
Zhang, B. 2023,
Rev. Mod. Phys., 95, 035005

\end{thebibliography}
\end{document}